\documentclass{appolb}
\usepackage{epsfig}

\newcommand{\Nc}{N_{\rm c}}

\newcommand{\lqcd}{\Lambda_{\rm QCD}}

\newcommand{\vp}{\vec{p}}
\newcommand{\vq}{\vec{q}}
\newcommand{\vk}{\vec{k}}

\newcommand{\la}{\langle}
\newcommand{\ra}{\rangle}

\newcommand{\pF}{p_{\rm F}}

\newcommand{\Np}{N_{\rm p}}
\newcommand{\rmd}{\mathrm{d}}
\newcommand{\rmi}{\mathrm{i}}
\newcommand{\rme}{\mathrm{e}}

\begin{document}
\title{Interweaving Chiral Spirals at finite quark density%
\thanks{Presented at 
``Three Days in Quarkyonic Island'', 19-21 May, 2011,
Wroclaw, Poland.}%
}
\author{Toru Kojo
\address{RIKEN BNL Research Center,
 Brookhaven National Laboratory, Upton, NY-11973, USA}
\address{Faculty of Physics, University of Bielefeld,
D-33501, Bielefeld, Germany}
}
\maketitle
\begin{abstract}
The interweaving chiral spirals (ICS), that is
defined as superposition of differently oriented chiral spirals,
is important for qualitative understandings
of the intermediate quark density region as well as
quantitative estimates of the Quarkyonic region.
We discuss how to construct the ICS,
taking the (2+1) dimensional Fermi system as an example.
We postulate that the presence of the ICS would delay the occurrence of
the chiral restoration as well as deconfinement phase transition,
by tempering the growth of quark fluctuations.
\end{abstract}
\PACS{12:38Aw, 11.30.Rd}

\section{Central Ideas: Qualitative Impacts of the ICS}
\begin{figure}[tb]
\vspace{-0.2cm}
\begin{center}
\scalebox{1.0}[0.9] {
\hspace{0.2cm}
  \includegraphics[scale=.20]{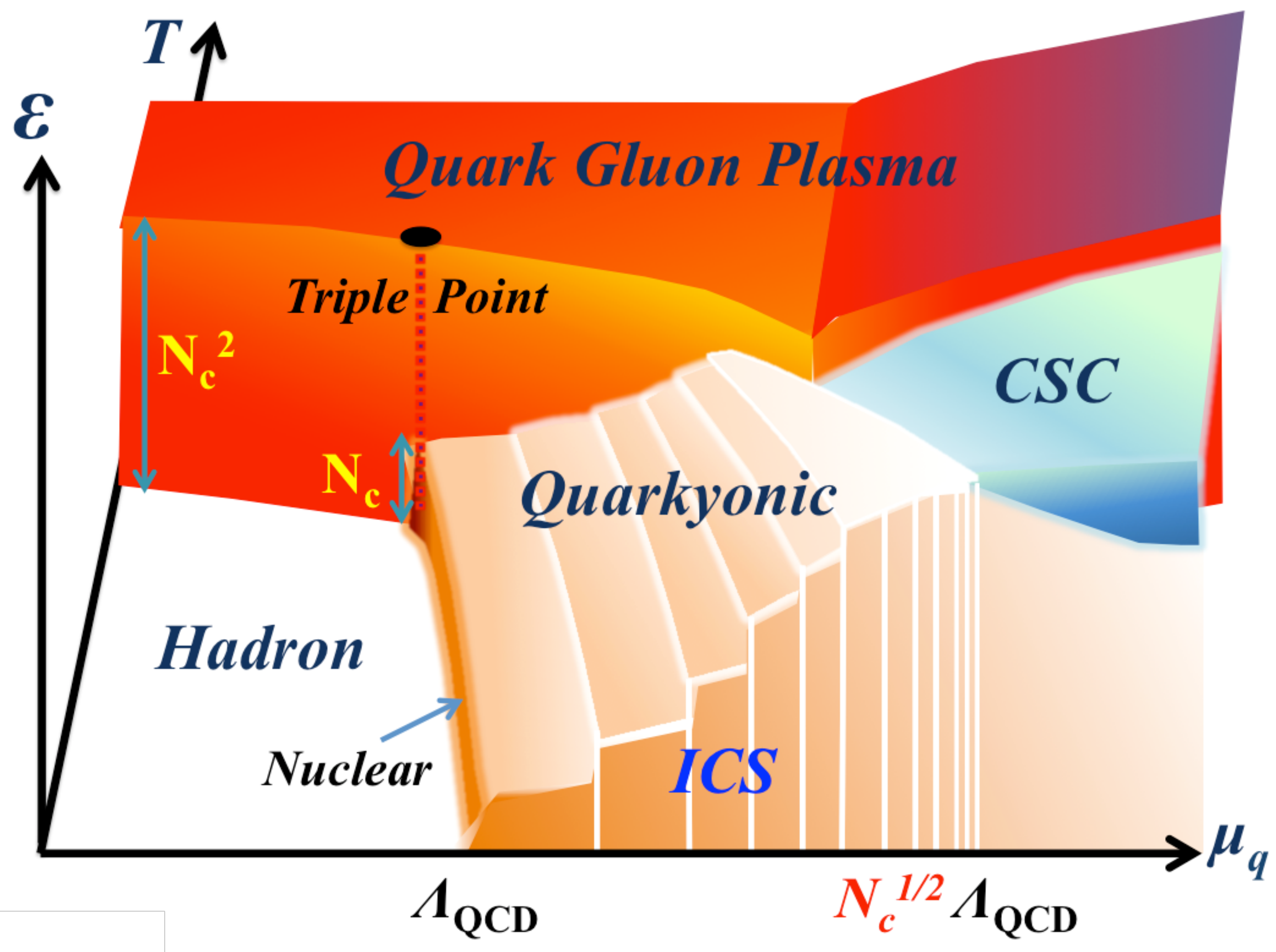} }
\end{center}
\vspace{-0.2cm}
\caption{A speculated QCD phase diagram.
}
\label{fig:Phase}
\vspace{-0.2cm}
\end{figure}

Recently,
it has been argued that there is a new state of QCD matter
at high baryon density and low to intermediate temperatures 
\cite{McLerran:2007qj}
(Fig.\ref{fig:Phase}).

This novel state is called Quarkyonic matter,
distinguished from
nuclear matter by its bulk quantities such as pressure.
The Fermi sea is mainly composed of quarks, not nucleons,
in the region a little bit above $\mu_q \sim M_N/\Nc \sim \lqcd$.
This is because after the emergence of nucleons,
a small change in $\mu_q$ rapidly enhances 
nucleon density, 
making strong short distance interactions among nucleons crucial. 
Thereby nucleons are 
not appropriate degrees of freedom to describe
the bulk part of the Fermi sea --
the quark picture is absolutely necessary 
to describe Quarkyonic matter.

Quarkyonic matter should be also distinguished from
conventional deconfined quark matter by its thermal
and Fermi surface excitations.
The excitations are confined, even after
quarks are released from nucleons.
A proper understanding of confined excitations is a basic starting point
for any discussions of phase structures, transport phenomena,
and for construction of the effective Lagrangian.

One of the most relevant observations in Ref. \cite{McLerran:2007qj}
is that the scale for the formation of the quark Fermi sea,
$\mu_q \sim \lqcd$, 
and that for the deconfinement of the excitations,
$\mu_q \sim \Nc^{1/(d-1)} \lqcd$ ($d$: spatial dimension), 
are conceptually different.
The latter may be estimated by comparing
quantum fluctuations of gluons with those of quarks near the Fermi surface.
This observation was tested for (1+1) dimensional QCD,
and it was argued that excitations are always confined\footnote{In spatial one dimension,
the phase space for quark fluctuations is always the same as vacuum case,
so the confining force is never modified.}
indendently of $\Nc$,
while the pressure is saturated by free quark contributions
\cite{Kojo:2011fh}.

Now let us ask what happens to chiral symmetry.
This issue is very important for qualitative understandings
of the intermediate density region 
as well as {\it quantitative} estimates for the Quarkyonic region.
Below we will argue that chiral symmetry is spontaneously {\it broken}
by inhomogeneous chiral spirals,
tempering the growth of the quark fluctuations at finite density.

Sometimes it is said that Quarkyonic matter is {\it defined} as
a chiral symmetric confined matter,
although chiral symmetry was {\it not} a primary issue in the original
proposal of Ref. \cite{McLerran:2007qj}.
If one sticks to this definition, it might be very difficult
to imagine the existence of 
Quarkyonic matter at $\Nc=3$.
Indeed, once the quarks near the Fermi surface become gapless,
it would largely enhance quark fluctuations
in addition to the phase space enhancement at finite density.
Recent results of the functional renormalization group application to
the PNJL model for not very high density \cite{Herbst:2010rf}
presumablly should be interpreted in this context.

\begin{figure}[tb]
\vspace{-0.2cm}
\begin{center}
\scalebox{1.0}[1.0] {
  \includegraphics[scale=.33]{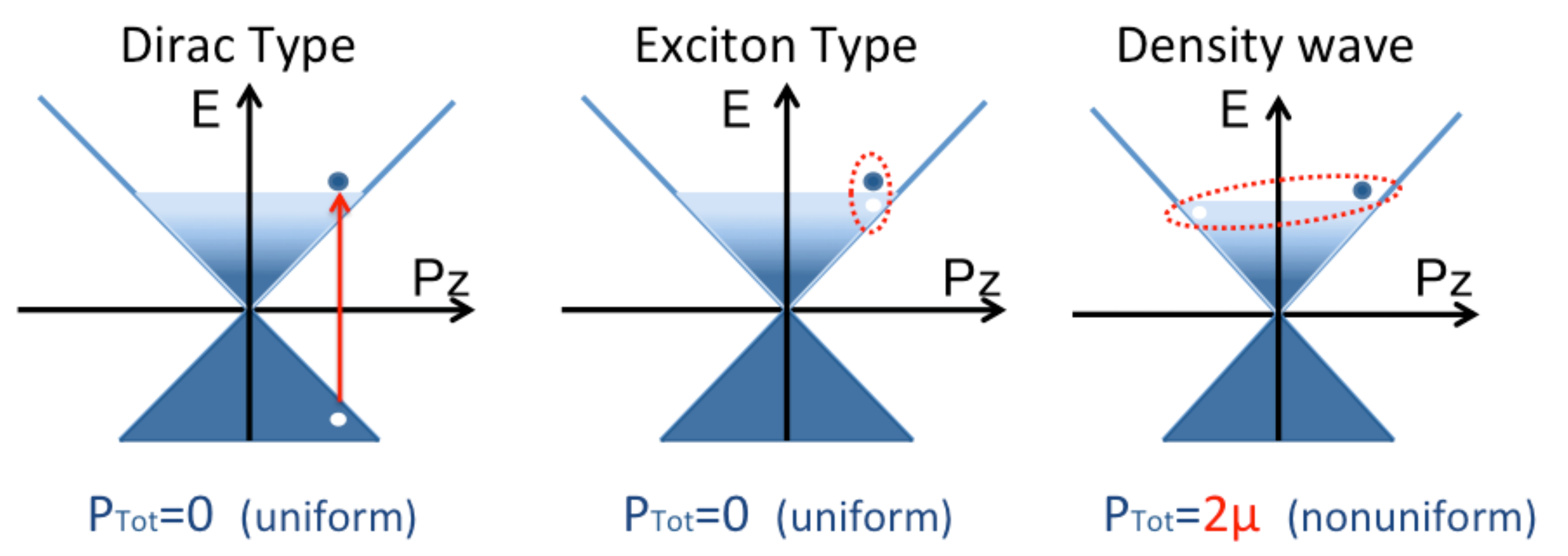} }
\end{center}
\vspace{-0.2cm}
\caption{Three types of chiral pairings.
}
\label{fig:pairing}
\vspace{-0.2cm}
\end{figure}

A general tendency of model analyses,
{\it without depending on whether models are confining or not},
suggests that the chiral restoration occurs shortly after
the formation of the quark Fermi sea, or
$p_F \sim \lqcd$ \cite{Glozman:2007tv}.
This trend may be understood by observing that
the creation of anti-quarks, which are
ingredients of the usual chiral condensate,
costs more energy for the larger quark Fermi sea.
This is so because the particle in the Dirac sea
must go above the Fermi surface
to avoid the Pauli-blocking
(Fig.\ref{fig:pairing}).

At finite density, however,
more proper ingredients of condensates
are particle-holes near the Fermi surface.
The excitations near the Fermi surface 
naturally have large momenta, $|\vp|\sim \mu_q$,
but they do not cost {\it additional} kinetic energy much,
compared to the energy before excitations.
Shown in Fig.\ref{fig:pairing} are the exciton and density wave
pairings.

In an exciton case,
the total momenum of a pair is $\sim 0$, 
but the relative momentum between a particle and a hole
is $\sim 2\mu_q$.
In a confining model,
such a pairing accompanies a large string,
costing large potential energy.

In a density wave case,
while the total momenum of a pair is $\sim 2\mu_q$,
a particle and a hole co-move without forming a large string.
Thus the density wave pairing should be energetically favored
compared to the exciton pairing.
(Arguments based on confining model picture
is useful but not indispensable, though. See below.)

Actually the chiral density wave solution can be always
interpreted as the chiral spirals.
A key observation is that
once we have a condensation of a pair moving to,
say, $+z$-direction, there is also a pair moving to $-z$ direction.
Mathmatically, one can project out fermion components
moving to $\pm z$ directions by operating the projection matrices
\cite{Kojo:2009ha},
\begin{equation}
\psi_\pm \equiv \frac{1 \pm \gamma_0 \gamma_z}{2} \psi ~.
\end{equation}
Then we have two types of the chiral condensates,
\begin{equation}
\la \bar{\psi}_- \psi_+ \ra \sim \Delta\, \rme^{2 \rmi \mu_q z} \, ,~~~
\la \bar{\psi}_+ \psi_- \ra \sim \Delta\, \rme^{-2 \rmi \mu_q z} \,,
\end{equation}
whose sum and difference give
\begin{equation}
\la \bar{\psi} \psi \ra \sim \Delta\, \cos(2 \mu_q z) \,,~~~
\la \bar{\psi} \rmi \gamma_0 \gamma_z\psi \ra 
\sim  \Delta\, \sin(2 \mu_q z) \,,
\end{equation}
with a fixed radius of $\Delta$ of the order $\lqcd^3$.
These condensates obviously break the chiral symmetry,
translational invariance, rotational invariance,
and the second condensate further breaks parity {\it locally}.

Here once again we emphasize that
in the presence of the Fermi sea,
large momenta naturally appear without costing much excitation energies,
so one should not be surprised at the
emergence of condensations with relatively large momenta.
Indeed, analyses of both non-confining 
\cite{Deryagin:1992rw,Shuster:1999tn,Nickel:2009ke,Rapp:2000zd}
and confining models \cite{Kojo:2009ha}
suggest that
the chiral spiral solution overtakes the homogeneous solution.

Rather a more nontrivial question is related to the fact that
the chiral spiral must have a particular orientation.
Let us ask: Can chiral pairs be formed in such a way to 
cover the entire Fermi surface, and
can differently oriented chiral spirals be interweaved
in a consistent way?

To answer to these questions will be very important 
for considerations about whether
the chiral symmetry breaking may survive
after taking the color superconductivity into account.
If only single chiral spiral in one particular direction were possible,
we could employ a less number of pairs for chiral condensations
than we do for diquark condensations.
If this would be the case,
the color superconducting phase would overtake
a single chiral spiral
\cite{Shuster:1999tn}.
But instead we shall suggest a possibility that
the ICS appears as far as the nonperturbative gluon exchange survives.

A possibility of the ICS has been qualitatively discussed
in a confining model \cite{Kojo:2010fe},
although whole aspects about the ICS were not fully explored
because of the technical difficulties in treating
the deep infrared region of the gluon exchange.
Actually, however, key aspects about the ICS
may be extracted without using the confinement,
and, in fact, are rather robust to the detailed
behaviors in the deep infrared region.

Below we shortly highlight
how to construct the ICS for the non-confining model
in (2+1) dimensions, 
together with the parametric estimates of several effects.
How to handle the corrections, 
relations with the previous works \cite{Nickel:2009ke,Rapp:2000zd}, 
etc.,
have been comprehensively discussed in a recent paper
\cite{Kojo:2011cn},
so an interested reader should consult it for details.

\section{How to Construct the ICS in (2+1) dimensions}
\begin{figure}[tb]
\vspace{0.0cm}
\begin{center}
\scalebox{1.0}[1.0] {
\hspace{0.0cm}
  \includegraphics[scale=.27]{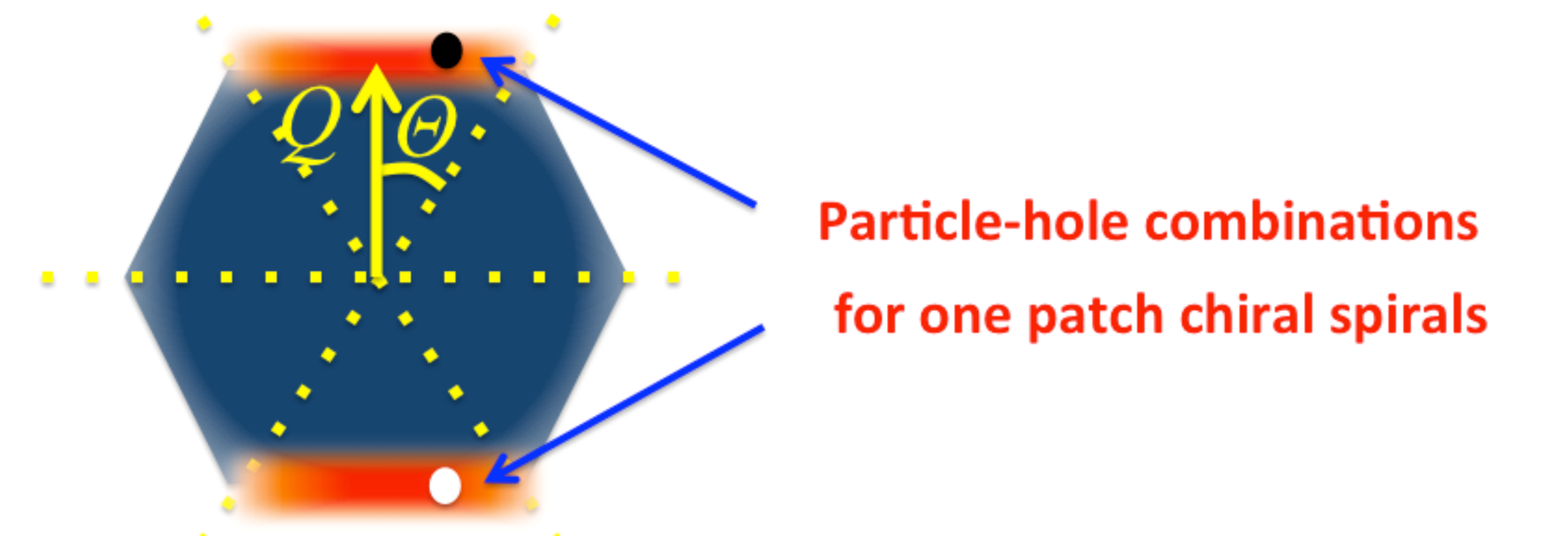} }
\end{center}
\caption{A 3-patches = 6-wedges case.
}
\label{fig:wedges}
\vspace{0.0cm}
\end{figure}

As an example of the ICS,
we consider a (2+1) dimensional Fermi sea\footnote{
Precisely speaking, in odd dimensional space-time,
the chirality is not defined, so the terminology
``chiral'' spirals may be a little bit misleading.
But the mechnism to generate spirals do not depend on
this fact.} at $T=0$,
purely because possible shapes of the Fermi surface are relatively simple.
An extension of our treatments to higher dimensional systems
are technically nontrivial but conceptually straightforward.

We first divide the Fermi sea into $2\Np$ wedges (Fig.\ref{fig:wedges}).
A wegde in one side of the Fermi sea and a wedge in the opposite side 
are regarded as a pair, and we call it one patch domain.
We denote
the height of each wedge as $Q$, and its open angle as 
$2\Theta = \pi/\Np$.
Each patch will generate a single chiral spiral.
The variables $Q$ and $\Theta$ are variational parameters
which will be optimized in such a way to minimize the total free energy.

Below we will use the canonical emsemble
with specifying quark number density or $p_F$,
because the presentation of ideas becomes simpler
than the grand canonical case.
Then the Fermi volume conservation can be used
to rewrite $Q$ as a function of $p_F$ and $\Theta$.
In the canonical emsemble,
our goal is to minimize the total energy\footnote{
At $T=0$, the minimization of the free energy
in the grand canonical emsemble is thermodynamically 
equivalent to the minimization of the total energy
in the canonical emsemble.}
by choosing the optimal value of $\Theta$.

\begin{figure}[t]
\vspace{0.0cm}
\begin{center}
\scalebox{1.0}[1.0] {
\vspace{-0.1cm}
  \includegraphics[scale=.25]{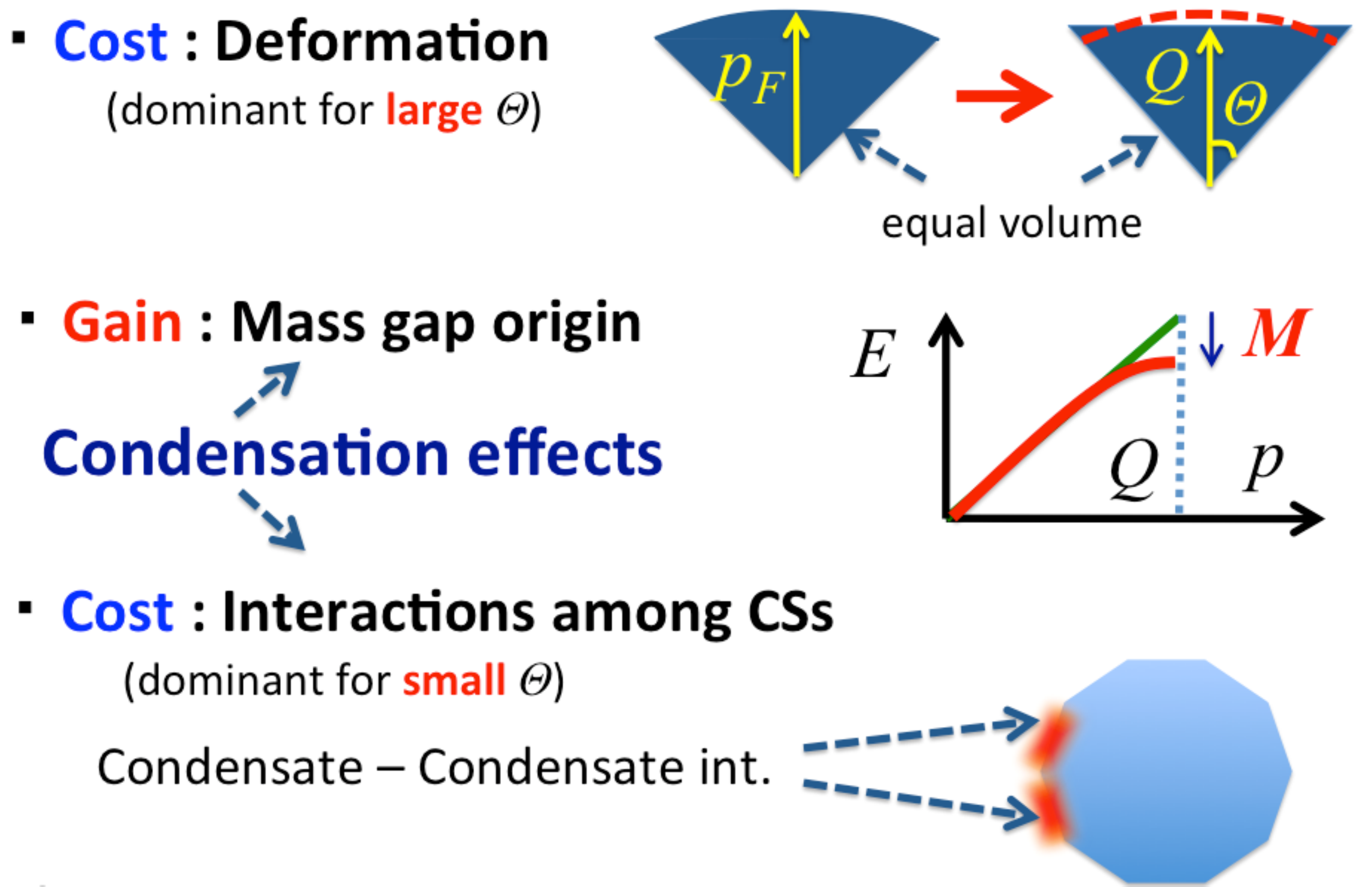} }
\end{center}
\caption{The energy gains and costs.
}
\label{fig:energy}
\vspace{-0.4cm}
\end{figure}

Essentially, the total energy and the shape of the Fermi surface
will be determined by balancing
the following energy costs and gains (Fig.\ref{fig:energy}):

\begin{itemize}
\item (i) The kinetic energy cost arising from the deformation of the
Fermi surface, which weakly depends upon the condensation effects.
This contibution becomes dominant 
for large $\Theta$.

\item (ii) The energy gain in a single patch from the condensation effects.
The condensation effects bend down the single particle dispersion.
Then particles occupy smaller energy orbits, 
reducing the total single particle energy.
In the following the quark mass gap will be denoted as $M$.
This contribution is less sensitive to the angle $\Theta$
compared to other contributions.

\item (iii) Coherent interactions among differently oriented chiral
spirals (we call it inter-patch interaction) which cost energy.
They originate from the condensation effects,
so become less important as condensates get smaller.
It becomes increasingly important
for small $\Theta$ due to the 
enhanced number of inter-patch interactions.
\end{itemize}

Below we shall give estimates for these contributions,
starting with the thereotically clean set up, 
that is,
the leading order (LO) of the $1/\Nc$ expansion and
the high density expansion in powers of $\lqcd/p_F$.
Although LO results are not directly applied to
the phenomenologically interesting region,
it is not difficult to specify which effects
will grow at lower density, 
and how results will be modified qualitatively.
Indeed, 
many features of the previous works \cite{Nickel:2009ke,Rapp:2000zd}, 
which have been numerically done for relatively lower density, 
can be understood from this analytic framework.

For explicit estimates,
it is indispensable to introduce models.
In particular, the momentum dependence of the interaction
strongly affects the estimate of the inter-patch interactions.
We will explain the features of the model,
then move to parametric estimates of several effects.
\subsection{A model}

\begin{figure}[tb]
\vspace{0.0cm}
\begin{center}
\scalebox{3.0}[1.0] {
  \includegraphics[scale=.10]{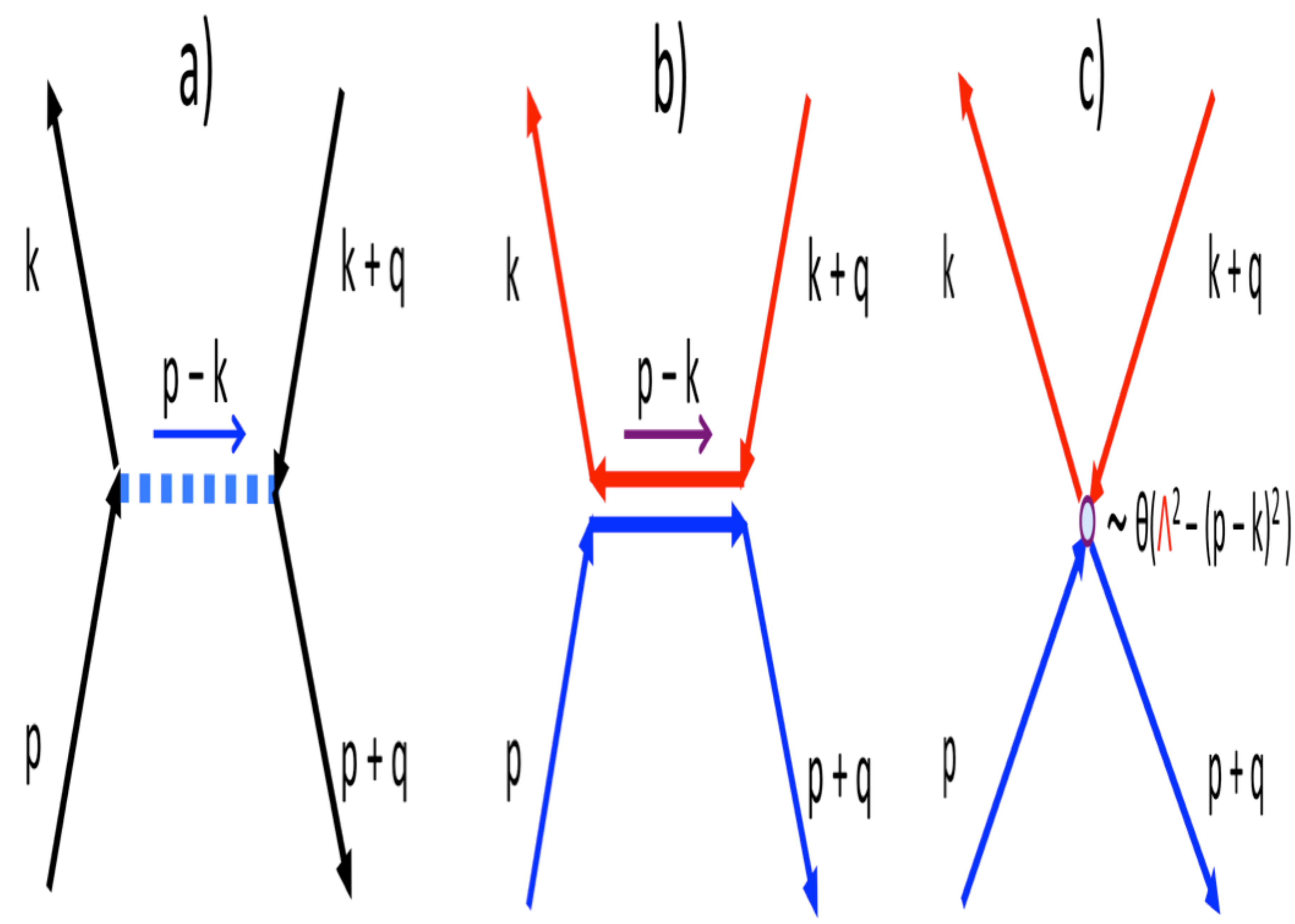} }
\end{center}
\vspace{-0.3cm}
\caption{
(a) The non-perturbative gluon exchange which is supposed to damp
  quickly in the UV region.
(b) The color line representation of the one-gluon exchange.
(c) Our effective four-Fermi interaction including form factor
  effects.
}
\label{figform}
\vspace{-0.2cm}
\end{figure}

In many studies of the intermediate density,
typically the 4-Fermi interaction has been used, 
with the ultraviolet (UV) cutoff on the quark momenta.
But this approach will be problematic when
the quark Fermi momentum $p_F$ becomes closer to the UV cutoff.
Thus we have to use alternative descriptions
at the intermediate density region.
Our model is the following non-local 4-Fermi interaction,
\begin{equation}
\hspace{-0.5cm}
\int \rmd^3x \, \big( \bar{\psi} \psi(x) \big)^2 \rightarrow
\int \rmd x_0 \! \int_{q,p,k}
\big( \bar{\psi}(\vp + \vq ) \psi(\vp) \big)
\big( \bar{\psi}(\vk ) \psi(\vk+ \vq ) \big) 
\,\theta_{p,k} \;,
\label{form2}
\end{equation}
where $\theta_{p,k} \equiv \theta\big( \lqcd^2 - (\vp - \vk)^2 \big)$,
and momentum integration is for $\vq, \vp, \vk$.
The large-$\Nc$ QCD is mimicked as follows.  The one-gluon exchange
including non-perturbative effects are shown in Fig.~\ref{figform}(a).
Its strength damps as the momentum transfer becomes large.  We roughly
take into account this property by introducing a step function,
$\theta\big( \lqcd^2 - (\vp-\vk)^2\big)$, keeping the interaction
strength constant.

In Fig.~\ref{figform}(b), we show the color line representation to
illustrate how the one-gluon exchange 
interaction should be contracted into a four-Fermi type interaction.
Taking into account features in Figs.~\ref{figform}(a) and
\ref{figform}(b), we arrive at a simple model described in
Eq.~(\ref{form2}) and Fig.~\ref{figform}(c).

The main consequence of our form factor treatments 
can be best seen in the Schwinger-Dyson or gap equations
(See Fig. \ref{figform}).
We illustrate it for zero density case.
After picking up a residue, we arrive at 
\begin{equation}
M(\vp) 
= \int \! \frac{ d \vk }{ (2\pi)^2 } ~
 \frac{M( \vk )} {2 \epsilon( \vk ) } ~
\theta_{p,k}\,.
~~~\left( \, \epsilon(\vk) = \sqrt{ M(\vk)^2 + \vk^2} \,\right)
\label{SDeq1}
\end{equation}
Note that because of $\theta_{p,k}$,
the contributions to the mass gap $M(\vp)$ comes from
the integral around $\vp$.
Putting in a different way,
a particle with $\vp$ is affected by the condensate
made of particles and anti-particles
with momenta close to $\vp$.
So quark-condensate interactions are
{\it local in momentum space}.
Also it is clear that at very large $|\vp|$,
the integrand quickly damps, leading to small $M(\vp)$.
The chiral restoration occurs for high energy excitations.

By applying this argument,
one can derive several conclusions about finite density.
The low energy excitations appear near the Fermi surface,
so that the chiral symmetry is violated
near the Fermi surface, 
but it is gradually restored in the region 
far from the Fermi surface.

Thanks to the locality in momentum space,
the inter-patch interactions among condensates
occur only if their momentum domains are close.
In the construction of the ICS,
this property is essential to restrict
interactions among
differently oriented chiral spirals
only near the patch boundaries.
If this locality is absent, 
inter-patch interactions would occur everywhere in the Fermi surface
destroying the chiral spirals one another, 
and would reduce the quark mass gap 
considerably \cite{Rapp:2000zd}.

Now we are ready to give qualitative estimates
of several contributions.

\begin{figure}[tb]
\vspace{0.0cm}
\begin{center}
\scalebox{0.6}[0.6] {
  \includegraphics[scale=.32]{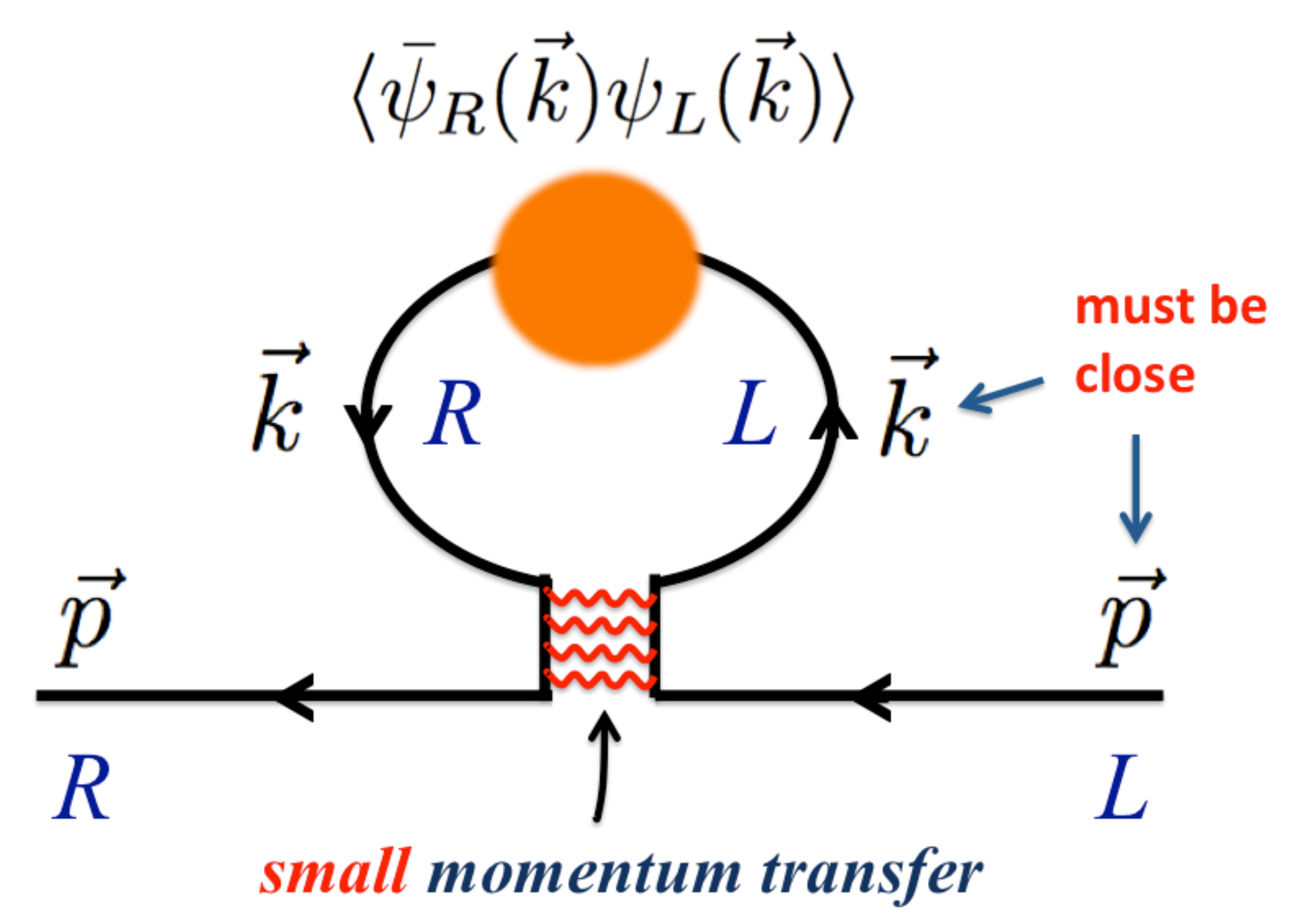} }
\scalebox{0.6}[0.6] {
  \includegraphics[scale=.32]{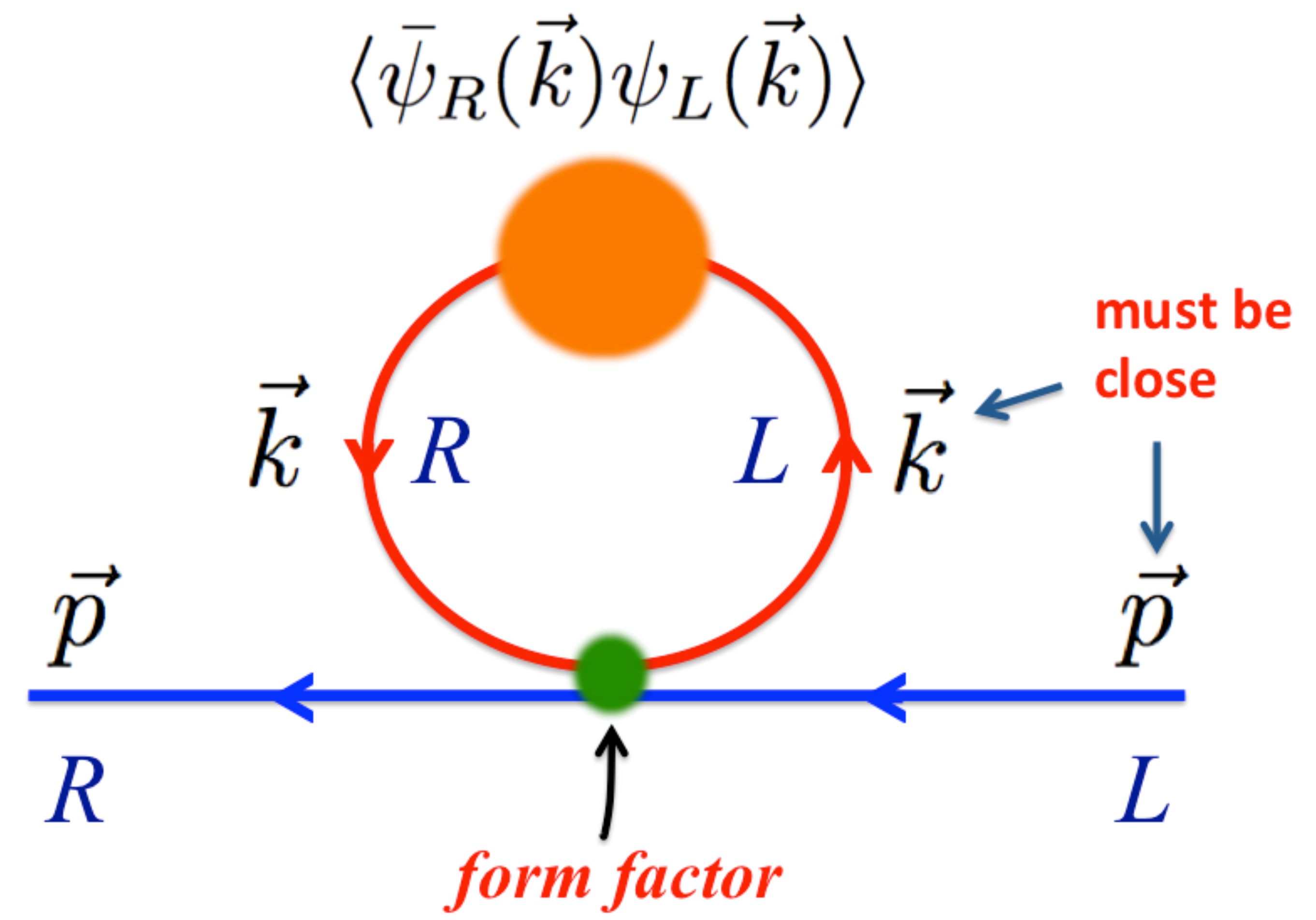} }
\end{center}
\vspace{-0.3cm}
\caption{
The leading
self-energy diagram at zero density.  
(Left) The diagram in terms of QCD dynamics
with the Rainbow Ladder approximation. 
(Right) The corresponding diagram in our model. 
}
\label{fig:SDeq}
\end{figure}
%
\subsection{The energy cost: Deformation energy}
The contribution (i) is rather easily estimated.
In one wedge, $Q$ is fixed by the Fermi volume conservation,
\begin{equation}
 \pF^2 \Theta = Q^2  \tan \Theta\,,
\end{equation}
and the difference between
the Deformed Fermi sea and the spherical Fermi sea can be
calculated as
\begin{equation}
\Delta{\mathcal E}_{ {\rm deform.} }(\Theta) 
\sim \Nc \cdot \pF^3\Theta^4 + O(\Theta^6) \,,
\end{equation}
where we have assumed $\Theta \ll 1$ and 
$M/p_F \ll 1$.
Interestingly, $\Theta^2$ term disappears
after the subtraction of the spherical Fermi sea.

\subsection{The energy gain: Single particle energy}
The solution of the gap equation 
gives the mass gap, $M \sim \lqcd$,
opened near the Fermi surface within distance of $\sim \lqcd$.
On the other hand, quarks outside of this domain
are not strongly affected by the condensates.
Thus the energy reduction after summing up
all patch contributions is
\begin{equation}
\Delta{\mathcal E}_{ {\rm cond.} }(\Theta) 
\, \sim \,
\Nc \cdot (- M) \times ( \lqcd \cdot p_F \tan \Theta) \times \Np
\, \sim \, - \Nc\lqcd^2 p_F \,,
\end{equation}
where $-M$ and
$\lqcd \cdot p_F \tan 2\Theta$
characterizes the bending down of
single particle energy,
a number of particles acquiring the mass gap
within a single patch, respectively.
A sum of all patch contributions
is approximately $\Theta$ independent,
and does not play a relevant role in the optimization of $\Theta$.

\subsection{The energy cost: Inter-patch interactions among condensates}
Since the single particle and the condensate interact 
only if their domains in momentum space are close one another,
so interactions between differently oriented chiral spirals occur
only near the patch boundaries.
The strength of the interaction is proportional to $M^2$.
Taking into account the phase space where interactions occur,
and counting a number of patch boudaries,
we estimate the energetic cost as
\begin{equation}
\Delta{\mathcal E}_{ {\rm int.} }(\Theta) 
\sim
\Nc\cdot M^2 \lqcd \times \Np 
\sim 
\Nc \lqcd^3/\Theta \,.
\end{equation}
Note that the contributions are proportional to $1/\Theta$,
so the creation of condensates are not favored for very small $\Theta$.
Actually, near the patch boundaries the quark mass gap becomes 
effectively smaller.

\subsection{The optimimal value of $\Theta$}
Now we can optimize $\Theta$ by
differentiating the total energy.
Since the single particle contributions
do not strongly depend on details of $\Theta$,
the optimal value of $\Theta$ is essentially
determined by balancing
the deformation energy and the inter-patch interaction energy:
\begin{equation}
\frac{1}{\Nc}
\frac{\partial}{\partial \Theta} 
\left( 
\Delta {\cal E}_{ {\rm deform.} } + \Delta {\cal E}_{ {\rm int.} } 
\right)
\, \sim \,
4 p_F^3 \Theta^3 - \frac{ \lqcd^3}{\Theta^2}
\, \sim \, 0 \,,
\end{equation}
which determines the optimal value of $\Theta$ as
\begin{equation}
\Theta \sim \left( \frac{\lqcd}{p_F} \right)^{3/5} \,.
\end{equation}
Therefore, a number of chiral spirals, $\Np$,
increases as density does.
Since $\Np$ is an integer,
the phase transition occurs discontinuously.

With this value of $\Theta$,
the deformation and interaction energies 
are $\sim p_F^{3/5} \lqcd^{12/5}$,
smaller than the single particle contribution,
$\sim -p_F \lqcd^2$.

\section{Summary}

We have argued why the ICS is potentially relevant, 
and shown
how to construct it by taking the (2+1) dimensional
Fermi system as an example.

The chiral spirals is a mechanism that can generate the quark mass gap
of $O(\lqcd)$ even after the formation of the quark Fermi sea.
We expect that the mass gap 
would temper the growth of the quark fluctuations
near the Fermi suface,
shifting chiral restoration and deconfinement lines
to higher temperature and density than those
predicted under the assumption of the homogeneous condensates.

We have argued the ICS only at $T=0$,
but for phenomenological applications
to the RHIC low energy scan or
future FAIR and NICA experiments,
the extension of the present results to
$T\neq 0$ are absolutely necessary.
We expect that some qualitative changes
occur at some temperature.
Such an extension will be presented in near future.

\section*{Acknowledgments}
The author thanks
the organizers for their kind hospitality.
Special thanks go to 
Y. Hidaka, K. Fukushima, L. McLerran, R.D. Pisarski,
and A.M. Tsvelik with whom
most of arguments presented here have been developed.
He also acknowledges
D. Blaschke,
E.J.~Ferrer, 
V.~Incera, 
J.M. Pawlowski,
A. Ohnishi,
G.~Torrieri
for discussions during the workshop,
and S.~Carignano and M.~Buballa
for explaining their studies on the chiral crystals 
before the publication.
He is supported by
RIKEN-BNL Research Center and
Humboldt foundation through
its Sofja Kovalevskaja program.



\begin{thebibliography}{00}
\bibitem{McLerran:2007qj}
  L.~McLerran, R.~D.~Pisarski,
  Nucl.\ Phys.\  A {\bf 796} (2007) 83.

\bibitem{Kojo:2011fh}
  T.~Kojo,
  [arXiv:1106.2187 [hep-ph]].

\bibitem{Herbst:2010rf}
  T.~K.~Herbst, J.~M.~Pawlowski and B.~J.~Schaefer,
  Phys.\ Lett.\  B {\bf 696} (2011) 58.

\bibitem{Glozman:2007tv}
  L.~Y.~Glozman, R.~F.~Wagenbrunn,
  Phys.\ Rev.\  D {\bf 77} (2008) 054027. 

\bibitem{Kojo:2009ha}
  T.~Kojo, Y.~Hidaka, L.~McLerran, R.~D.~Pisarski,
  Nucl.\ Phys.\  A {\bf 843} (2010) 37.

\bibitem{Deryagin:1992rw}
  D.~V.~Deryagin, D.~Y.~Grigoriev, V.~A.~Rubakov,
  Int.\ J.\ Mod.\ Phys.\  A {\bf 7} (1992) 659.
%
\bibitem{Shuster:1999tn}
  E.~Shuster, D.~T.~Son,
  Nucl.\ Phys.\  B {\bf 573} (2000) 434; 
  B.~Y.~Park, M.~Rho, A.~Wirzba, I.~Zahed,
  Phys.\ Rev.\  D {\bf 62} (2000) 034015.

\bibitem{Nickel:2009ke}
  D.~Nickel,
  Phys.\ Rev.\ Lett.\  {\bf 103} (2009) 072301;
  Phys.\ Rev.\  D {\bf 80} (2009) 074025;
  S.~Carignano, D.~Nickel, M.~Buballa,
  Phys.\ Rev.\  D {\bf 82} (2010) 054009.
\bibitem{Rapp:2000zd}
  R.~Rapp, E.~V.~Shuryak, I.~Zahed,
  Phys.\ Rev.\  D {\bf 63} (2001) 034008.
\bibitem{Kojo:2010fe}
  T.~Kojo, R.~D.~Pisarski, A.~M.~Tsvelik,
  Phys.\ Rev.\  D {\bf 82} (2010) 074015. 

\bibitem{Kojo:2011cn}
  T.~Kojo, Y.~Hidaka, K.~Fukushima, L.~McLerran, R.~D.~Pisarski,
    [arXiv:1107.2124 [hep-ph]].
\end{thebibliography}
\end{document}